\documentclass[12pt]{article}
\usepackage{amssymb,latexsym}
\usepackage{amsmath}
\usepackage{multirow}
\usepackage{amsfonts}

\makeatletter \@addtoreset{equation}{section} \makeatother

\begin{document}

\newcommand{\nn}{\nonumber}
\newcommand{\miso}{\frac{1}{2}}
\def\beq{\begin{equation}}
\def\eeq{\end{equation}}
\def\bea{\begin{eqnarray}}
\def\eea{\end{eqnarray}}
\def\mc{\mathcal}
\newcommand{\m}{\mathbf}
\newcommand{\fet}{\frac{1}{3}}
\newcommand{\fdt}{\frac{2}{3}}
\newcommand{\ftt}{\frac{4}{3}}
\def\w{\wedge}
\def\olra{\overleftrightarrow}

\begin{titlepage}

\thispagestyle{empty}

\vspace{20pt}

\begin{center}

{ \Large{\bf Reducing the Heterotic Supergravity \\on
nearly-K\"{a}hler coset spaces}}

\vspace{35pt}

{\bf A.~Chatzistavrakidis}$^{1,2}$,  {\bf P.~Manousselis}$^{2}$,
{\bf and} {\bf G.~Zoupanos}$^2$ \vspace{20pt}

$^1$ {\it Institute of Nuclear Physics,\\
NCSR  Demokritos,\\
GR-15310 Athens, Greece}\\
 \vspace{5pt}

$^2${\it Physics Department, National Technical University of Athens, \\
GR-15780 Zografou Campus, Athens, Greece} \\
\vspace{5pt}

\vspace{5pt}

\

Email: {\tt cthan@mail.ntua.gr,  pman@central.ntua.gr,
george.zoupanos@cern.ch}

\vspace{35pt}

$\mathbf{{Abstract}}$
\end{center}

We study the dimensional reduction of the ${\cal N}=1$,
ten-dimensional Heterotic Supergravity to four dimensions, at
leading order in $\alpha'$, when the internal space is a nearly-K\"{a}hler
manifold. Nearly-K\"{a}hler manifolds in six dimensions are all the non-symmetric coset spaces and a group manifold. Here we reduce the theory using as internal manifolds
the three six-dimensional non-symmetric coset spaces, omitting the case of the group manifold in the prospect of obtaining chiral fermions when the gauge fields will be included.  We determine the
effective actions for these cases, which turn out to describe ${\cal N}=1$
four-dimensional supergravities of the no-scale type and we study the various possibilities concerning their vacuum. 

\end{titlepage}

\newpage

\baselineskip 6 mm

\section{Introduction}

Supergravity theories have been studied extensively over the past
thirty years. In particular, exploring the possibility that
superstring theories describe the real world, the task of providing
a suitable compactification which would lead to a realistic
four-dimensional theory has been pursued in many diverse ways.

The early attempts to reduce such theories made extensive use of
Calabi-Yau (CY) manifolds, i.e. manifolds with $SU(3)$ holonomy
\cite{Candelas:1985en}. A reduction and truncation procedure has
been developed in refs. \cite{Witten:1985xb} and
\cite{Derendinger:1985kk}, where the internal space is not specified
but the general characteristics of CY manifolds are kept. However,
there exist some problems with the use of CY in the reduction
procedure due to their complicated geometry. Among others their
metric is not known explicitly and their Euler characteristic is too
large to accommodate an acceptable number of fermion generations.
Moreover, in CY compactifications the resulting low-energy field
theory in four dimensions contains a number of massless chiral
fields, characteristic of the internal geometry, known as moduli.
These fields correspond to flat directions of the effective
potential and therefore their values are left undetermined. Since
these values specify the masses and couplings of the
four-dimensional theory, the theory has limited predictive power.

In the context of flux compactifications the recent developments
have led to the study of a wider class of internal spaces, called
manifolds with $SU(3)$-structure, that contains CYs. The general case of $SU(3)$-structures is
of special interest since the ''local Lorentz'' (structure) group
$SO(6)$ of the internal space can be reduced down to $SU(3)$ in a
way that there exists a nowhere-vanishing globally-defined spinor.
In the case of CY manifolds this spinor is covariantly constant with respect to the Levi-Civita connection, while it can be constant with respect to a connection with torsion in the general case.
The latter condition allows for a wider class of internal spaces, such as
nearly-K\"{a}hler and half-flat manifolds. The Heterotic String
theory has been recently studied in this general context in refs.\cite{Gurrieri:2007jg} and \cite{Benmachiche:2008ma}. Six-dimensional nearly-K\"{a}hler manifolds are all the non-symmetric six-dimensional coset spaces plus the group manifold $SU(2)\times SU(2)$ and they have been identified as supersymmetric
solutions in the case of type II theories (see e.g.
\cite{Lust:2004ig}-\cite{House:2005yc}). In the studies of
compactification of the Heterotic Supergravity the use of non-symmetric coset
spaces was introduced in \cite{Govindarajan:1986kb}, and recently
developed further in \cite{Lopes Cardoso:2002hd}-\cite{Manousselis:2005xa}.
Particularly, in \cite{Manousselis:2005xa} it was shown that
supersymmetric compactifications of the Heterotic String theory of
the form $AdS_4\times S/R$ exist when background fluxes and general
condensates are present. In addition, effective theories have been
constructed in \cite{House:2005yc}, \cite{KashaniPoor:2007tr},
\cite{Caviezel:2008ik} in the case of type II supergravity.

Here we would like to discuss the dimensional reduction of the
Heterotic String at leading order in $\alpha'$ in the case where the
internal manifold admits a nearly-K\"{a}hler structure. In section 2 we provide a brief
reminder of the Heterotic Supergravity and discuss the basics of
manifolds with $SU(3)$-structure. In section 3 we present the
general reduction procedure that we follow and determine the
resulting four-dimensional Lagrangian. In section 4 we apply the
previously found results in the case of six-dimensional non-symmetric coset spaces (i.e. in all nearly-K\"{a}hler manifolds, omitting the case of the group manifold since it cannot lead to chiral fermions in four dimensions) and we discuss the supergravity
description from the four-dimensional point of view. Section 5
contains a discussion on the inclusion of gauge fields in our
framework. Finally, our conclusions appear in section 6.

\section{General Framework}

\subsection{Heterotic Supergravity}

In this section we briefly review the field content and the
Lagrangian of the Heterotic Supergravity in order to fix our
notation and conventions.

The field content of the Heterotic Supergravity consists of the
${\cal N}=1$, $D=10$ supergravity multiplet, which accommodates the
fields $g_{MN}$, $\psi_{M}$, $B_{MN}$, $\lambda$ and $\varphi$ (i.e.
the graviton, the gravitino which is a Rarita-Schwinger field, the
two-form potential, the dilatino which is a Majorana-Weyl spinor,
and the dilaton which is a scalar). Capital Latin letters denote
here ten-dimensional indices.

The corresponding Lagrangian of the ten-dimensional ${\cal N}=1$
Heterotic Supergravity in the Einstein frame can be written as
\cite{Chapline:1982ww} \beq \mc{L}=\mc{L}_b+\mc{L}_f+\mc{L}_{int},
\eeq where the different sectors of the theory, ignoring the gauge
fields and the gaugini at lowest order, are {\footnote{Here we use
differential form notation for the kinetic terms of the bosons,
which will prove to be useful in the course of the reduction.}} \bea
\hat{e}^{-1}\mc{L}_{b}&=&
-\frac{1}{2\hat{\kappa}^2}(\hat{R}\hat{*}\mathbf{1}+ \frac{1}{2}
e^{- \hat{\phi}}\hat{H}_{(3)}\wedge \hat{*}\hat{H}_{(3)}+\frac{1}{2}
d\hat{\phi}\wedge
                \hat{*} d\hat{\phi}),\\
 \hat{e}^{-1}\mc{L}_{f}&=&-\miso{\hat{\bar\psi}}_M \hat{\Gamma}^{MNP}
            D_N\hat{\psi}_P
                 - \miso{\hat{\bar\lambda}}\hat{\Gamma}^M D_M \hat{\lambda},\\
 \hat{e}^{-1}\mc{L}_{int}&=& -\miso{\hat{\bar\psi}}_M \hat{\Gamma}^N \hat{\Gamma}^M
                            \hat{\lambda}
                                \partial_N\hat{\phi}+e^{-\hat{\phi}/2}\hat{H}_{PQR}\biggl(\hat{\bar{\psi}}_M\hat{\Gamma}^{MPQRN}\hat{\psi}_N
                                    +6{\hat{\bar\psi}^P}\hat{\Gamma}^Q\hat{\psi}^R\nn\\&-& \sqrt{2}\bar{\psi}_M\hat{\Gamma}^{PQR}\hat{\Gamma}^M \hat{\lambda})\biggl)
                                 +\mbox{four-fermion terms},
                                               \eea
where we have placed hats in all the ten-dimensional fields to
distinguish them from their four-dimensional counterparts which will
appear after the reduction. The three-form $\hat{H}_{(3)}$ is the
field strength for the $B$-field, namely $\hat{H}=d\hat{B}$. The
gamma matrices are the generators of the ten-dimensional Clifford
algebra, hence we place hats on them too, while those with more than
one indices denote antisymmetric products of $\Gamma$s. Also,
$\hat{\kappa}$ is the gravitational coupling constant in ten
dimensions, with dimensions (mass)$^4$; $\hat{e}$ is the determinant
of the metric and $\hat\ast$ is the Hodge star operator in ten
dimensions. Let us mention that the gravitational constant is
defined as $\hat{\kappa}^2=8\pi G_N$, where $G_N$ is the Newton
constant. As such its relation to the Planck mass is $\hat{\kappa}=
\frac{1}{m_{Pl}}$, since the Planck mass is
$m_{Pl}=\frac{1}{\sqrt{{8\pi G_N}}}$.

\subsection{Manifolds with $SU(3)$-structure }

CY manifolds were proposed as internal spaces for
compactifications in view of the requirement that a four-dimensional
${\cal N}=1$ supersymmetry is preserved. Namely they admit a single
globally defined spinor, which is covariantly constant with respect
to the (torsionless) Levi-Civita connection. However, there exists a
larger class of manifolds for which the spinor is covariantly
constant with respect to a connection with torsion. These are called
manifolds with $SU(3)$-structure and clearly CY manifolds
are a subclass in the category of $SU(3)$-structure manifolds.

In particular, in order to define a nowhere-vanishing spinor on a
six-dimensional manifold one has to reduce the structure group
$SO(6)$. The simplest procedure one can follow is to reduce this
group to $SU(3)$, since then the decomposition of the spinor of
$SO(6)$ reads $\mathbf{4}=\mathbf{3}+\mathbf{1}$ and the spinor we
are looking for is the singlet, let us call it $\eta$. Then, we can
use $\eta$ to define the $SU(3)$-structure forms, which are a
real two-form $J$ and a complex three-form $\Omega$ defined as \bea J_{mn} &=& \mp i\eta_{\pm}^{\dag}\gamma_{mn}\eta_{\pm}, \nn\\
    \Omega_{mnp} &=& \eta_-^{\dag}\gamma_{mnp}\eta_+,\nn\\
    \Omega^*_{mnp} &=& -\eta_+^{\dag}\gamma_{mnp}\eta_-, \eea where the signs denote the chirality of the spinor and the normalization is $\eta_{\pm}^{\dag}\eta_{\pm}=1$.
    These forms are globally-defined and non-vanishing and they are
subject to the following compatibility conditions \bea J\w J\w J &=&
\frac{3}{4}i\Omega\w\Omega^*, \nn\\ J\w\Omega&=&0. \eea Moreover,
they are not closed forms but instead they satisfy
\bea\label{strfrms} dJ &=& \frac{3}{4}i({\cal
W}_1\Omega^*-{\cal
W}^*_1\Omega)+{\cal W}_4\w J+{\cal W}_3, \nn\\
    d\Omega &=& {\cal W}_1J\w J+{\cal W}_2\w J +{\cal W}_5^*\w
    \Omega. \eea
The expressions (\ref{strfrms}) define the five intrinsic torsion
classes, which are a zero-form ${\cal W}_1$, a two-form ${\cal
W}_2$, a three-form ${\cal W}_3$ and two one-forms ${\cal W}_4$ and
${\cal W}_5$. These classes completely characterize the intrinsic
torsion of the manifold. Note that the classes ${\cal W}_1$ and
${\cal W}_2$ can be decomposed in real and imaginary parts as ${\cal
W}_1={\cal W}_1^++{\cal W}_1^-$ and similarly for ${\cal W}_2$.

The classes ${\cal W}_1$ and ${\cal W}_2$ are vanishing when the
manifold is complex and furthermore a K\"{a}hler manifold has
vanishing ${\cal W}_3$ and ${\cal W}_4$. A Calabi-Yau manifold has
all the torsion classes equal to zero. An interesting class of
$SU(3)$-structure manifolds are called nearly-K\"{a}hler manifolds.
In this case all the torsion classes but ${\cal W}_1$ are vanishing.
This suggests that the manifold is not K\"{a}hler and not even
complex. A complete list of other classes of $SU(3)$-structure
manifolds can be found in \cite{Grana:2005jc}.

Manifolds with $SU(3)$-structure in general
%\cite{Gurrieri:2007jg},\cite{Benmachiche:2008ma}
and nearly-K\"{a}hler manifolds in particular
%\cite{Micu:2004tz},\cite{KashaniPoor:2007tr}
have attracted a lot of interest in flux compactifications over the
last years. Here we are interested in six-dimensional nearly-K\"{a}hler manifolds, which have been classified in \cite{Butruille:2006}. They are the three non-symmetric six-dimensional coset spaces, namely
$G_2/SU(3),Sp_4/(SU(2)\times U(1))_{non-max}$
and $SU(3)/U(1)\times U(1)$, plus the group manifold $SU(2)\times
SU(2)$. It is therefore interesting to perform an explicit reduction
over these spaces{\footnote{We shall omit in our discussion the case
of the group manifold and treat only the coset spaces in the
prospect of obtaining chiral fermions when the gauge sector will be
added.}} and determine the resulting effective actions, a task which
we shall perform in the forthcoming sections.

\section{Reduction procedure}

In the present section we focus on the bosonic part of the Heterotic
Supergravity Lagrangian and perform its reduction from ten to four
dimensions over the coset spaces S/R. Since the K\"{a}hler potential
and the superpotential of the four-dimensional theory can be
obtained from the bosonic part, this procedure will be sufficient to
find the supergravity description in four dimensions.

In order to reduce the theory we need ansatze for the bosons, namely
for the metric, the dilaton and the $B$-field. Starting with the
metric, our ansatz reads \begin{equation}\label{gransatz} d\hat{s}_{(10)}^2 =
e^{2\alpha\varphi(x)} \eta_{mn} e^{m} e^{n} +
e^{2\beta\varphi(x)}\gamma_{ab}(x)e^{a}e^{b},
\end{equation}
where $e^{2\alpha\varphi(x)}\eta_{mn}$ is the four-dimensional metric and $e^{2\beta\varphi(x)}\gamma_{ab}(x)$ is the internal metric, while $e^m$ are the one-forms of the orthonormal basis in four dimensions and $e^a$ are the left-invariant one-forms on the coset space. In this ansatz we included exponentials which rescale the
metric components. This is always needed in order to obtain an
action without any prefactor for the Einstein-Hilbert part. We shall
see that we need to specify the values of $\alpha$ and $\beta$ in
order to fulfil this requirement.

We note that the ansatz (\ref{gransatz}) is dictated by two further
requirements. Firstly, the metric is required to be $S$-invariant.
Secondly, the requirement of consistency of our reduction enforces
the vanishing of the Kaluza-Klein (KK) fields and allows only the
scalar fluctuations
\cite{Chatzistavrakidis:2007by}-\cite{Chatzistavrakidis:2007pp}. In
particular, tackling the consistency problem, direct calculations
lead to the result that when KK gauge fields take values in the
maximal isometry group of the coset space, $S\times N(R)/R$
{\footnote{Here, $N(R)$ denotes the normalizer of $R$ in $S$, which
is defined as $N = \{ s \in S, \ \ \  sRs^{-1} \subset R \}.$ Note
that since $R$ is normal in $N(R)$ the quotient $N(R)/R$ is a
group.}}, the lower-dimensional theory is, in general, inconsistent
with the original one. Full consistency of the effective Lagrangian
and field equations with the higher-dimensional theory is guaranteed
when the KK gauge fields are ($N(R)/R$)-valued \cite{Coquereaux:1986zf}. However, when the
condition $rankS=rankR$ holds the group $N(R)/R$ is trivial. This is
the case for the spaces we consider and therefore the KK gauge
fields vanish. Finally, the  part of the internal
metric $\gamma_{ab}(x)$ without the exponential has to be unimodular.

Following the standard procedure (see e.g. \cite{Cvetic:2003jy}) for reducing the Einstein-Hilbert action in the case
of a coset space  and choosing
$\alpha = -\frac{\sqrt{3}}{4},\beta = -\frac{\alpha}{3}$, we find
that the corresponding part of the reduced Lagrangian reads
\begin{eqnarray}\label{redlang1}
\mathcal{L}=-\frac{1}{2\kappa^2}\biggl(R*\mathbf{1}-P_{ab}\wedge*
P_{ab}+\frac{1}{2}d\varphi\wedge* d\varphi\biggl)-V,
\end{eqnarray} with the potential V having the form \begin{equation}
V=-\frac{1}{8\kappa^2}e^{2(\alpha-\beta)\varphi}(\gamma_{ab}\gamma^{cd}\gamma^{ef}f^{a}_{
\ ce}f^{b}_{ \
 df}+2\gamma^{ab}f^{c}_{ \ da}f^{d}_{ \ cb}+4\gamma^{ab}f_{iac}f^{ic}_{b}
 )*\mathbf{1},
\end{equation} where the index $i$ runs in $R$ and $\kappa=\frac{\hat{\kappa}}{vol_6}$ is the gravitational coupling constant in four dimensions. In the reduced Lagrangian the fields $P_{ab}$ are defined as
\beq P_{ab} = \frac{1}{2}\biggl[(\Phi^{-1})^{c}_{a}d\Phi^{b}_{c} +
(\Phi^{-1})^{c}_{b}d\Phi^{a}_{c}\biggl], \eeq with $\Phi^a_b$
defined through the relation \begin{equation} \gamma_{cd} =
\delta_{ab} \Phi^{a}_{c} \Phi^{b}_{d}.
\end{equation} As such, $\Phi$ is a matrix of unit determinant,
generically containing scalar fields other than $\varphi$, and hence
there exists a set $(\Phi^{-1})^{b}_{a}$ of fields satisfying
\begin{equation}
(\Phi^{-1})^{c}_{a}(\Phi^{-1})^{d}_{b}\gamma_{cd} = \delta_{ab}.
\end{equation}
The corresponding kinetic term in (\ref{redlang1}) will provide the
kinetic terms for the extra scalars apart from $\varphi$, which are
generically needed to parametrize the most general $S$-invariant
metric and appear through the unimodular metric $\gamma_{ab}(x)$.

As far as the higher-dimensional dilaton is concerned, it is trivially
reduced by $\hat{\phi}(x,y) = \phi(x)$, since it is already a scalar
in ten dimensions. This leads to a kinetic term
$-\frac{1}{4\kappa^2} d\phi\w*d\phi$ in the reduced Lagrangian.

Finally, concerning the three-form sector of the theory we expand
the $B$-field on $S$-invariant forms of the coset space; namely our
ansatz reads \begin{equation}\label{twoform} \hat{B}= B(x) +
b^i(x)\omega_i(y),
\end{equation} where the index i counts the number of $S$-invariant
two-forms. Then it is straightforward to see that the
higher-dimensional three-form $\hat{H}=d\hat{B}$ can be written in
terms of four-dimensional fields as \beq\label{hexp} \hat{H}=
dB+db^i\w\omega_i+b^id\omega_i. \eeq Let us note here that unlike
the case of CY compactifications, where the expansion forms are
harmonic and hence closed, here we expand in forms that are not
closed and thus an extra term appears in eq.(\ref{hexp}). Note in addition that at
the order we are working it is straightforward to see that
$d\hat{H}=0$ and therefore our ansatz (\ref{twoform}) solves the
Bianchi identity as it should.

In order to determine the reduced Lagrangian we need to dualize the
expression (\ref{hexp}) with respect to the ten-dimensional Hodge
star operator. Then we find \bea\label{hdual} \hat{*}_{10}\hat{H} =
e^{-6\alpha\varphi}*_4dB\w vol_6 +
e^{-2\alpha\varphi-4\beta\varphi}*_4db^i\w*_6\omega_i+e^{-6\beta\varphi}b^ivol_4\w\ast_6d\omega_i.
\eea Moreover, the determinant of the metric is
$\hat{e}=e^{2\alpha\varphi}$.

Using the expressions (\ref{hexp}) and (\ref{hdual}) in the
corresponding term in the Lagrangian we find that the reduced
Lagrangian for this sector becomes \bea\label{hlang}  {\cal L}=
-\frac{1}{2\kappa^2}e^{-\phi}\biggl[\miso
e^{-4\alpha\varphi}d\theta\w\ast
d\theta+\frac{m}{2}e^{-4\beta\varphi}db^i\w\ast db^i\nn\\+\miso
e^{4\alpha\varphi}(n_1(b^i)^2+n_2\epsilon_{ij}b^ib^j)vol_4\biggl]\w
vol_6, \eea where $\theta$ is the pseudoscalar obtained by duality
transformation on $dB$, while $m$, $n_1$ and $n_2$ are fixed
constants defined by \bea\label{mn} \omega_i\w\ast\omega_j = m
\delta _{ij}vol_6,~~~ d\omega_i\w*d\omega_j =
(n_1\delta_{ij}+n_2\epsilon_{ij})vol_6.  \eea

Let us conclude this section by adding some comments concerning the possibility of
including a background flux for the ten-dimensional field strength
$\hat{H}$. Since fluxes can be included as additional sources with
indices purely in the internal manifold, we have two
three-forms at our disposal, $\rho_1$ and $\rho_2$, as we shall
see in the concrete examples of the following section. Therefore one
could in principle include in $\hat{H}$ a term proportional to either $\rho_1$ or
$\rho_2$ or both. Note that for the spaces we use it always holds
that the structure form $\Omega$ is proportional to a complex linear
combination of these three-forms and particularly to
$\rho_2+i\rho_1$. However, we can check that the exterior derivative of any invariant
two-form is proportional to $\rho_2$. This means that the inclusion
of a term proportional to $\rho_2$ is redundant since it can always
be absorbed in the definition of the scalar fields $b^i$. This property is intimately connected with the fact that in the
nearly-K\"{a}hler limit the only non-vanishing torsion class is
${\cal W}_1$. As we discussed in section 2.2, ${\cal W}_1$ can be
split in real and imaginary parts. In our cases the real part is
always vanishing and there exists only an imaginary part for this
torsion class. Therefore the remaining possibility is to introduce a
flux proportional to $\rho_1$. However, this is a non-closed form.
Then, the addition of such a term would mean that the Bianchi
identity would fail to hold, since $d\hat{H}$ would not vanish
anymore. This in turn means that at this level no background flux
can be added. The situation certainly could change when gauge fields are
taken into account.

As a final remark let us note that in refs.\cite{House:2005yc} and \cite{KashaniPoor:2007tr} the suitable basis of expansion forms for nearly-K\"{a}hler manifolds has been specified and actually coincides with our basis of $S$-invariant forms.

\section{Examples}

In this section we specialize the previous discussion in the case of
non-symmetric six-dimensional coset spaces, namely $G_2/SU(3)$,
$Sp_4/(SU(2)\times U(1))_{non-max}$ and $SU(3)/U(1)\times U(1)$. We
determine the potential in four dimensions and we find the
corresponding supergravity description by defining the appropriate
K\"{a}hler potential and superpotential. All these spaces admit a
nearly-K\"{a}hler structure and therefore our results are to be
compared to the results of refs.\cite{Gurrieri:2007jg} and
\cite{Benmachiche:2008ma}. Indeed, as we shall see, our models are
realizations of the formalism of the articles \cite{Gurrieri:2007jg}
and \cite{Benmachiche:2008ma}. In particular, the superpotentials we
find can be retrieved through the Heterotic Gukov formula found in
\cite{Gurrieri:2007jg} (see also \cite{Lopes Cardoso:2003af}).

\

\textbf{\underline{Geometry and SU(3)-structure}}

\

\textbf{\textit{G${}_2$/SU(3)}:} According to ref.\cite{MuellerHoissen:1987cq} this
manifold has one $G_2$-invariant two-form given by
\beq\omega_1=e^{12}-e^{34}-e^{56},\eeq and two $G_2$-invariant three-forms expressed as \bea\rho_1&=&e^{136}+e^{145}-e^{235}+e^{246},\nn\\ \rho_2
&=&e^{135}-e^{146}+e^{236}+e^{245},\eea in terms of the coset indices $1\dots 6$ which correspond to the complement of $SU(3)$ in $G_2$. On the other hand invariant one-forms do not exist. The
invariant forms of the coset space are intimately connected to its
$SU(3)$-structure forms $J$ and $\Omega$. Indeed $J$ and $\Omega$ are given by \bea J &=&
R\omega_1,\nn\\ \Omega &=& \sqrt{R^3}(\rho_2+i\rho_1), \eea where R
is the radius of the space and we can immediately deduce that \beq dJ =
-\sqrt{3}R\rho_2 = -\sqrt{3}Re(\Omega). \eeq Then, from the first
equation in (\ref{strfrms}) we can read that the torsion classes
${\cal W}_3$ and ${\cal W}_4$ are vanishing as expected, while it is
straightforward to see that \beq {\cal W}_1= -\frac{2i}{\sqrt{3R}}.  \eeq

Finally, determining that
$d\Omega=\frac{8i}{\sqrt{3}}R(e^{1234}+e^{1256}-e^{3456})$ we can
see that the second equation in (\ref{strfrms}) is consistently
satisfied with ${\cal W}_1$ as above and ${\cal W}_2={\cal W}_5=0$.
Thus we find that for this coset space the only non-vanishing
torsion class is ${\cal W}_1$, which means that it naturally admits
a nearly-K\"{a}hler structure without any further conditions.

 \

\textbf{\textit{Sp${}_4$/(SU(2)${}\times$U(1))${}_{non-max}$}:} Here there exist two
$Sp_4$-invariant two-forms given by \bea\omega_1 &=&
e^{12}+e^{56},\nn\\\omega_2 &=& e^{34}\eea and two three-forms expressed as
\bea\rho_1 &=& e^{136}-e^{145}+e^{235}+e^{246},\nn\\\rho_2 &=&
e^{135}+e^{146}-e^{236}+e^{245}.\eea The indices $1\dots 6$ are coset indices corresponding to the complement of $SU(2)\times U(1)$ in $Sp_4$. As in the previous case
invariant one-forms do not exist. The structure forms are given by \begin{eqnarray}
J &=& -R_1\omega_1+R_2\omega_2,\nn\\ \Omega &=& \sqrt{R_1^{2} R_2
}(\rho_2+i\rho_1),
\end{eqnarray} with $R_1$ and $R_2$ the radii of the space. The
non-vanishing torsion classes in this case are \begin{eqnarray}
{\cal W}_1 &=& -\frac{2i}{3} \frac{2R_1+R_2}{\sqrt{R_1^2 R_2}}, \\
{\cal W}_2 &=& -\frac{4i}{3} \frac{1}{\sqrt{R_1^2 R_2}}\left[
R_1(R_1 -R_2) e^{12} - 2R_2(R_2 - R_1)e^{34} + R_1(R_1-R_2)e^{56}
\right]
\end{eqnarray} and it is obvious that the space has a
nearly-K\"{a}hler limit when the condition $R_1=R_2$ is satisfied.

\

\textbf{\textit{SU(3)/U(1)${}\times$U(1)}:} The coset space $SU(3)/U(1)\times
U(1)$ has three $SU(3)$-invariant two-forms given by
\bea\omega_1&=&e^{12},\nn\\\omega_2 &=& e^{45},\nn\\\omega_3
&=&e^{67},\eea two invariant three-forms expressed as
\bea\rho_1&=&e^{147}-e^{156}+e^{246}+e^{257},\nn\\\rho_2
&=&e^{146}+e^{157}-e^{247}+e^{256},\eea while invariant one-forms do not exist.
In this case the indices 3 and 8 correspond to the two $U(1)$s and
the rest are coset indices corresponding to the complement of $U(1)\times U(1)$ in $SU(3)$. The forms which specify the
$SU(3)$-structure are
\begin{eqnarray} J &=& -R_1 \omega_1 + R_2 \omega_2 -
R_3\omega_3,\nn\\ \Omega &=& \sqrt{ R_1R_2R_3}(\rho_2+i\rho_1),
\end{eqnarray} where the three radii of the space are involved,
while the torsion classes are \begin{eqnarray}
{\cal W}_1 &=& -\frac{2i}{3} \frac{R_1+R_2+R_3}{\sqrt{R_1R_2R_3}}, \\
{\cal W}_2 &=& -\frac{4i}{3} \frac{1}{\sqrt{R_1R_2R_3}}[
R_1(2R_1 -R_2-R_3) e^{12} - R_2(2R_2 - R_1-R_3)e^{34} \nn\\
&+& R_3(2R_3-R_1-R_2)e^{56}].
\end{eqnarray} Again it is straightforward to see that under the condition
of equal radii this space admits a nearly-K\"{a}hler structure.

\

\

\textbf{\underline{Supergravity description in four dimensions}}

\

\textbf{\textit{G${}_2$/SU(3)}:} For $G_2/SU(3)$ the most general
$G_2$-invariant metric is given by \beq
g_{ab}=e^{2\beta\varphi}\delta_{ab},\eeq namely there is only one
scale and one scalar field $\varphi$ parametrizing the internal
metric. Thus, using the fact that $\gamma_{ab}=\delta_{ab}$ as well
as the structure constants of this coset space
\cite{Govindarajan:1986kb}, we easily find that the four dimensional
potential in this case is \beq\label{pot1}
V=-\frac{1}{2\kappa^2}(10e^{\frac{8\alpha}{3}\varphi}-6e^{-\phi+4\alpha\varphi}b^2).
\eeq Note that we found $n_1=12$ for the coefficient $n_1$ appearing in eq.(\ref{hlang}) in the present case. In order to bring the
reduced Lagrangian in the standard four-dimensional supergravity
form we define the complex superfields, consisting of all the scalar
moduli, \bea S&=& e^{\phi_0}+i\lambda,\nn\\
T&=&e^{-\varphi_0/\sqrt{3}}+ib. \eea Here for convenience we have
redefined two of the moduli as \bea
\phi_0&=&\miso(-\phi-4\alpha\varphi),\nn\\
\varphi_0&=&\miso(-\varphi+4\alpha\phi). \eea
Then we claim that the K\"{a}hler potential and the superpotential have the form \bea\label{kpot}
K&=&\frac{1}{\kappa^2}\biggl[-ln(S+\overline{S})-3ln(T+\overline{T})\biggl],\\\label{wpot}W&=&4\sqrt{3}T.
\eea Indeed we can easily verify that the four-dimensional
potential (\ref{pot1}) results from the supergravity expression
\beq\label{spot}
V(\Phi,\bar{\Phi})=\frac{1}{\kappa^4}e^{\kappa^2K}\big(K^{i\bar{j}}\frac{DW}{D\Phi^i}\frac{D\overline{W}}{D\bar{\Phi}^{\bar{j}}}-3\kappa^2W\overline{W}\big),
\eeq where the complex superfields are collectively denoted by
$\Phi$ and the derivatives involved are the K\"{a}hler covariant
derivatives \beq \frac{DW}{D\Phi^i}=\frac{\partial
W}{\partial\Phi^i}+\frac{\partial K}{\partial\Phi^i}W,\eeq for $K$ and $W$ given in eqs.(\ref{kpot}) and (\ref{wpot}) respectively. Also one
can check that the kinetic terms are exactly retrieved as \beq
-\frac{1}{\kappa^2}K_{i\bar{j}}d\Phi^i\w\ast d\bar{\Phi}^{\bar{j}}
\eeq with the same K\"{a}hler potential, as required by
supergravity.

 \

\textbf{\textit{Sp${}_4$/(SU(2)${}\times$U(1))${}_{non-max}$}:} This coset space
admits two independent scales therefore we need to parametrize the metric
by an extra scalar field, say $\chi$. Then the metric can be written
as \beq g_{ab}=e^{2\beta \varphi}\mbox{diag}(e^{2\gamma
\chi},e^{2\gamma \chi},e^{-4\gamma \chi},e^{-4\gamma
\chi},e^{2\gamma \chi},e^{2\gamma \chi}). \eeq To ensure the correct
kinetic term for the new scalar field we choose
$\gamma^2=\frac{1}{24}$. Also, there are now two scalars from the
$B$-field, $b_1$ and $b_2$. In the same spirit as in the previous
case we find that the four-dimensional supergravity description is
obtained by defining the K\"{a}hler potential and superpotential as follows
\bea
K&=&\frac{1}{\kappa^2}\biggl[-ln(S+\overline{S})-ln[(T_1+\overline{T}_1)^2(T_2+\overline{T}_2)]\biggl],\\W&=&-2T_1+T_2,
\eea    where the complex superfields are defined as \bea S&=&
e^{\phi_0}+i\lambda,\nn\\
T_1&=&e^{-\varphi_0/\sqrt{2}}+ib_1,\nn\\T_2&=&e^{-\chi_0}+ib_2. \eea
Note that the redefinitions \bea \phi_0 &=&
-\miso(\phi+4\alpha\varphi),\\
\varphi_0&=&-\frac{\sqrt{2}}{2}(\phi-4\alpha\varphi+4\gamma\chi),\\
\chi_0&=&-\miso(\phi-\frac{2\alpha}{3}\varphi-4\gamma\chi)\eea are
needed in order to ensure the consistency of the previous
expressions. The nearly-K\"{a}hler limit corresponds to the case
$T_1=T_2$.

 \

\textbf{\textit{SU(3)/U(1)${}\times$U(1)}:} Here there exist in principle three independent
scales, namely there exist three scalars parametrizing the metric
fluctuations $\varphi,\chi,\psi$ and on the other hand there exist
three fields $b_1,b_2,b_3$ in this case. We write for the metric
\beq g_{ab}=e^{2\beta
\varphi}\mbox{diag}(e^{2(\gamma\chi+\delta\psi)},e^{2(\gamma\chi
+\delta\psi)},e^{2(\gamma\chi-\delta\psi)},e^{2(\gamma\chi-\delta\psi)},e^{-4\gamma\chi},e^{-4\gamma
\chi}). \eeq Correctly normalized kinetic terms for $\chi$ and
$\psi$ are obtained with the choices $\gamma^2=\frac{1}{24}$ and
$\delta^2=\frac{1}{8}$.

The same logic as in the previous cases leads in the present one to the K\"{a}hler potential \beq
K=\frac{1}{\kappa^2}\biggl[-ln(S+\overline{S})-ln[(T_1+\overline{T}_1)(T_2+\overline{T}_2)(T_3+\overline{T}_3)]\biggl]
\eeq and the superpotential \beq W=-T_1+T_2-T_3. \eeq The scalar
superfields are now defined as \bea S&=&
e^{\phi_0}+i\lambda,\nn\\
T_1&=&e^{-\varphi_0}+ib_1,\\T_2&=&e^{-\chi_0}+ib_2,\\T_3&=&e^{-\psi_0}+ib_3\eea
with the redefinitions \bea \phi_0 &=& -\miso(\phi+4\alpha\varphi) \\
\varphi_0 &=& -\miso(\phi-\frac{4\alpha}{3}\varphi+4\gamma\chi+4\delta\psi) \\
 \chi_0 &=& -\miso(\phi-\frac{4\alpha}{3}\varphi+4\gamma\chi-4\delta\psi)
 \\ \psi_0 &=& -\miso(\phi-\frac{4\alpha}{3}\varphi-8\gamma\chi).
\eea The nearly-K\"{a}hler limit is obtained again when
$T_1=T_2=T_3$.

\

\textbf{\underline{Vacua}}

\

In the preceding examples we found that the reduction from ten
dimensions to four at leading order in $\alpha'$ leads to ${\cal
N}=1$ supergravities in four dimensions. Let us now study the possible vacua of the four-dimensional theory.

Requiring existence of a supersymmetric vacuum, the F-equations \beq\label{a2} \frac{DW}{D\Phi^i}=0 \eeq have to be satisfied. Then from eq.(\ref{spot}) we deduce that the vacuum energy in four dimensions is negative semidefinite. Thus as long as
supersymmetry remains unbroken it is impossible to find a de Sitter
vacuum \cite{de Wit:1986xg}. Moreover, the possibility to have
Minkowski vacuum suggests that the potential in eq.(\ref{spot})
vanishes in the vacuum, which in turn means that in addition the condition \beq\label{a1} W=0 \eeq has to be satisfied.
Unfortunately, the set of equations (\ref{a2}) and (\ref{a1}) cannot be satisfied in general. For example in
the
case of $G_2/SU(3)$ the above requirements lead to the equations  \bea W&=&0, \\ D_SW &=& -\frac{W}{S+\overline{S}}=0, \\
D_TW&=&4\sqrt{3}-\frac{3W}{T+\overline{T}}=0, \eea which are obviously inconsistent. Therefore either (i) supersymmetry is preserved and the four-dimensional space is not Minkowski or (ii) supersymmetry is broken and the four-dimensional space can be Minkowski. The same result
holds in the other two cases.
Since we deal with a theory of gravity it is natural to impose that the cosmological constant vanishes, at least at tree level, and elaborate further the option (ii) above. In the case of $G_2/SU(3)$ inspecting eq.(\ref{pot1}) we observe that we can tune the potential to vanish by imposing appropriate relations among the vacuum expectation values of the four-dimensional scalar fields. In particular, if we impose \beq 5e^{\frac{8\alpha}{3}<\varphi>}=3e^{-<\phi>+4\alpha<\varphi>}<b>^2 \eeq
or equivalently, in terms of the redefined fields, \beq 5e^{-\frac{2}{\sqrt{3}}<\varphi_0>}=3<b>^2, \eeq it is straightforward to see that in the vacuum the potential vanishes. Clearly this vacuum is not supersymmetric. Indeed, supersymmetry is spontaneously broken (due to the super-Higgs effect) and the gravitino obtains a mass \beq m_{3/2}=e^{-<G>/2}, \eeq where the function $G$ is defined as \beq G=K+ln(|W|^2). \eeq The graviton of course remains massless and therefore appears a splitting in the supergravity multiplet.

An obvious suggestion in order to go further in the examination of the possible vacua is to take into account the gauge fields, which have been neglected in the present examination, and also to include background
fluxes \cite{Strominger:1986uh}. We comment on that in the following section.

\section{Inclusion of gauge fields}

The results of our analysis indicate the necessity of including
gauge fields in our models and thus working at first order in
$\alpha'$. This is done by coupling the ${\cal N}=1$ supergravity
multiplet to an ${\cal N}=1$ vector supermultiplet consisting of the
gauge fields $\hat{A}_M$ and their superpartners, the gaugini
$\hat{\chi}$. It is well-known that the cancelation of anomalies
allows only the gauge groups $E_8\times E_8$ and $SO(32)$
\cite{Green:1984sg} and therefore the gauge fields and the gaugini transform
in the adjoint representation of one of these gauge groups.

The bosonic part of the ten-dimensional Lagrangian contains the term
\beq \hat{e}^{-1}{\cal
L}_{b,gauge}=-\frac{\alpha'}{2\hat{\kappa}^2}e^{-\hat{\phi}/2}Tr(\hat{F}\w\hat{\ast}\hat{F}),
\eeq where $\hat{F}$ is the field strength of the gauge field
$\hat{A}_M$. Moreover, the three-form $\hat{H}$ is now given by \beq
\hat{H}= d\hat{B}-\alpha'(\hat{\omega}_{YM}-\hat{\omega}_{L}), \eeq
where the Chern-Simons forms are defined as usual \beq
\hat{\omega}_{YM} =
Tr(\hat{F}\wedge\hat{A}-\frac{1}{3}\hat{A}\wedge\hat{A}\wedge\hat{A}),
\eeq \beq \hat{\omega}_L=Tr(\hat{\theta}\wedge
d\hat{\theta}+\frac{2}{3}\hat{\theta}\w\hat{\theta}\w\hat{\theta}),
\eeq where we denote the spin-connection with $\hat{\theta}$. These
two corrections are necessary to cancel completely the anomalies
(gauge, gravitational and mixed) of ${\cal N}=1$, $D=10$
supergravity coupled to Yang-Mills.

In order to dimensionally reduce the full bosonic
Lagrangian of the Heterotic String we need an ansatz for the gauge
fields. An interesting possibility emerges when one uses the Coset
Space Dimensional Reduction (CSDR) scheme \cite{Forgacs:1979zs},\cite{Kapetanakis:1992hf}. The
CSDR is based on the requirement that the gauge fields are not
invariant under the isometries of the coset space but their
transformation is compensated by a gauge transformation. This
requirement restricts the possible ansatze for the gauge fields
\cite{Chatzistavrakidis:2007by}. Using the CSDR scheme we can
benefit from several results that have been accomplished over the
years. Among those we refer the possibility to find four-dimensional chiral theories \cite{Chapline:1982wy}, as well as softly broken supersymmetric Lagrangians \cite{Manousselis:2001re}.

Concerning the possibility to obtain realistic models,
interesting four-dimensional GUTs have been found in refs \cite{Kapetanakis:1992hf}-\cite{Manousselis:2001re} resulting from ten-dimensional ${\cal N}=1$
supersymmetric $E_8$ gauge theories using CSDR. Moreover a rather complete
classification of the theories obtained starting from the same ten-dimensional theory by CSDR followed by a subsequent application of the Wilson flux breaking mechanism
has been recently given in \cite{Douzas:2008va}. Obviously, more possibilities to obtain realistic models might appear when one includes in the study the $E_8\times E_8$ as initial gauge group and relaxes the condition that the discrete symmetries are freely-acting, as it was assumed in the above study.

Finally, note that the reduction of the gauge
sector of the ten-dimensional theory will lead to an enhanced
potential. Then the K\"{a}hler potential and the superpotential of
the four-dimensional theory will receive $\alpha'$ corrections. The
possible vacua of the extended models have to be explored again. We plan to report on this work in a
forthcoming publication.

\section{Discussion and Conclusions}

Here we have explicitly reduced the Heterotic
Supergravity from ten dimensions to four at leading order in
$\alpha'$, i.e. ignoring the gauge fields, using six-dimensional nearly-K\"{a}hler manifolds as
internal spaces. We have examined three specific models based on
the three six-dimensional non-symmetric coset spaces admitting a nearly-K\"{a}hler structure and we have
determined the resulting four-dimensional effective actions.

From our results we observe that in all three cases the K\"{a}hler potential and the superpotential of the resulting
four-dimensional supergravities have the same structure and they differ only on
the number of scalar moduli appearing in each case. In particular,
the volume of the internal space is parametrized by one scalar field
for $G_2/SU(3)$, while in the cases of $Sp_4/(SU(2\times
U(1))_{non-max}$ and $SU(3)/U(1)\times U(1)$ the volume depends on
two and three scalar fields respectively. In addition, the
scalar fields emerging from the internal components of the $B$-field
are one, two and three respectively in the three examples we
have studied. It is worth noting that the structure of the K\"{a}hler
potential is exactly the one appearing in the no-scale models of
supergravity \cite{Lahanas:1986uc}. No-scale supergravity is an
effective theory exhibiting very interesting features such as that it leads to a
vanishing cosmological constant at the classical level, dynamical
determination of all mass scales in terms of the Planck scale and
potentially realistic low-energy phenomenology\footnote{The no-scale
structure as a low-energy limit of superstring theories has been
derived before in e.g. \cite{Witten:1985xb},\cite{Derendinger:1985kk}.}.

Concerning the vacuum of these models, as long as
supersymmetry remains unbroken it is impossible to find a de Sitter
vacuum. On the other hand, obtaining a
Minkowski vacuum would mean that the cosmological constant in four
dimensions vanishes. Examining the conditions which have
to be satisfied in the case of unbroken supersymmetry and vanishing vacuum energy we find that
there is no such solution in any of our models at this order in $\alpha'$. On the other hand, imposing the vanishing of the vacuum energy leads to a non-supersymmetric vacuum with supersymmetry spontaneously broken.
In order to enrich our models and look for realistic phenomenology, we are naturally led
to work on the next order in $\alpha'$ and include
gauge fields and non-vanishing fluxes.
Then the hope is that working in the context of the CSDR we shall be able to find interesting
supergravity GUTs in four dimensions with an appropriate number of
fermion generations and soft breaking of supersymmetry. This work is currently in
progress.

\vspace{3cm} % delate this line

\textbf{Acknowledgements} The authors would like to thank P.
Aschieri, G.L. Cardoso, A. Kehagias, C. Kounnas, G. Koutsoumbas, D.
L\"{u}st and D. Tsimpis for useful discussions. This work is
supported by the NTUA programme for basic research ''Karatheodoris''
and the European Union's RTN programme under contract
MRTN-CT-2006-035505.
%\clearpage % delate this line

\end{document}